\documentclass[12pt]{article}
\usepackage{epsfig}

\newcommand{\Z}{{Z \!\!\! Z}}
\newcommand{\beqn}{\begin{eqnarray}}
\newcommand{\eeqn}{\end{eqnarray}}
\newcommand{\eq}[1]{(\ref{#1})}
\newcommand{\const}{{\mathrm{const}}\,}
\newcommand{\dd}{\mbox{d}}

\newcommand{\dual}{\mbox{}^{\ast}}
\newcommand{\phot}{{\mathrm{phot}}}
\newcommand{\mon}{{\mathrm{mon}}}

\newcommand{\itep}
{~\vspace{-1.5cm}
\begin{flushright}
{\large LU-ITP 2002/013}\\
{\large KANAZAWA-02-16}\\
{\large ITEP-LAT-2002-07}\\
{\large RCNP-Th02014}
\end{flushright}
\vspace{1.0cm}}

\begin{document}
\baselineskip=14pt
\begin{center}

\itep

{\large\bf  String Breaking and Monopoles: \\
a Case Study in the 3D Abelian Higgs Model}

\vskip 1.0cm {\large
M.~N.~Chernodub$^{a,b}$, E.--M.~Ilgenfritz$^c$ and A.~Schiller$^d$}\\

\vspace{.4cm}

{ \it
$^a$ ITEP, B. Cheremushkinskaya 25, Moscow, 117259, Russia

\vspace{0.3cm}

$^b$ ITP, Kanazawa University, Kanazawa 920-1192, Japan

\vspace{0.3cm}

$^c$ RCNP, Osaka University, Osaka 567-0047, Japan

\vspace{0.3cm}

$^d$ Institut f\"ur Theoretische Physik and NTZ, Universit\"at
Leipzig,\\ D-04109 Leipzig, Germany
}

\end{center}

\begin{abstract}
We study the breaking of the string spanned between test charges
in the three dimensional Abelian Higgs model with compact gauge
field and fundamentally charged Higgs field at zero temperature.
In agreement with current expectations we demonstrate that string
breaking is associated with pairing of monopoles. However, the
string breaking is not accompanied by an ordinary phase transition.
\end{abstract}

\section{Introduction}
\label{sec:introduction}

The lattice Abelian Higgs model with compact gauge field (cAHM) in
three dimensions is of a broad interest both for high energy
physics~\cite{FradkinShenker,EinhornSavit} and condensed matter
physics~\cite{NagaosaLee,KleinertNogueiraSudbo,RecentSudbo} --
where it was suggested to describe high--$T_c$
superconductors and strongly correlated electron systems. Nowadays, it
has even entered the physics of cognitive
networks~\cite{FujitaMatsui}.

Due to compactness of the gauge field the model possesses Abelian
monopoles which are instanton--like excitations in three
space--time dimensions. The Abelian monopoles are able -- if they
are in the plasma state -- to accomplish confinement of
electrically charged particles. This is well known from cQED$_3$
where opposite
charged particles are bound by a linear potential~\cite{Polyakov}. The
confinement is arranged by monopoles forming an opposite charged
double sheet along the surface spanned by the trajectories of the
external test charges. This surface is usually considered as the
world surface of a string. Due to screening, the free energy
increases only proportional to the area of the surface such that an area law
for the Wilson loop emerges.

However, if dynamical matter fields in the same representation as
the external test charges are added to the confining theory, linear
confinement may be lost. This should be so, irrespective
whether the dynamical matter field is fermionic (the quarks in
QCD) or bosonic (the Higgs particle in our case).
The string breaking phenomenon has been
extensively studied in non--Abelian gauge theories with matter
fields~\cite{StringBreaking}
or with test charges in the adjoint
representation~\cite{AdjointStringBreaking}.
Here we want to investigate string
breaking in cAHM$_3$ with a $q=1$ charged Higgs field, a model
whose permanently confining counterpart, cQED$_3$, is well
understood. The general, intuitive picture says in the present
case that the string breaks {\it because} of Higgs particle pairs
popping up out of the vacuum at a definite inter--particle
separation between the external, infinitely heavy test charges.
Thus, the physical state corresponding to a {\it broken string}
would consist of two heavy--light mesonic states plus some number
of light-light Higgs pairs.

In order to destroy the linearly rising potential in cAHM$_3$, the
coupling between the Higgs field and the gauge field must be
sufficiently strong. One might be tempted to associate the string
breaking with a phase transition between confinement and Higgs
phases. Indeed, Ref.~\cite{KleinertNogueiraSudbo} proposes to
associate the string breaking with a
Berezinsky-Kosterlitz--Thouless type transition. In this paper we
demonstrate that and how the expected string breaking happens in a
part of the phase diagram where a first or second order phase
transition can definitely be excluded.

Abelian monopoles play the crucial role in the dual
superconductivity scenario~\cite{DualSuperconductivity} of
confinement in QCD. There, the monopole degrees of freedom need to
be defined with the help of Abelian
projections~\cite{AbelianProjections} (see, {\it e.g.}
reviews~\cite{Review}). The condensed magnetic currents were shown
to make a dominant contribution to the string tension between
quarks, in pure $SU(2)$ gauge theory~\cite{AbelianDominance} as
well as in $SU(3)$ gluodynamics and also in full
finite--temperature QCD with $N_f=2$ flavors of dynamical
quarks~\cite{MonopolesQCD}. Moreover, in full QCD with dynamical
quarks the contribution of Abelian monopoles to the heavy--quark
potential QCD shows the property of string
breaking~\cite{StringBreakingQCD}. The breaking of the adjoint
string in pure gluodynamics as well as the breaking of the
fundamental string in full QCD can both be described within the
Abelian projection formalism~\cite{SuzukiChernodub}. The
back--reaction of the dynamical fermions on the gauge field should
modify the dynamics of monopoles in such a way that this dynamics
incorporates the above qualitative
picture~\cite{StringBreakingQCD}.

Therefore, guided by the analogy to QCD, we focus our interest in
the present paper on the monopole degrees of freedom in compact
AHM in three dimensions under the influence of a scalar matter
field. We would like to elucidate the changing role of monopoles
under the particular aspect of string breaking.  As in QCD, the
string tension in this model is exclusively due to monopoles.
Therefore one can expect that monopoles also encode the
back--reaction of the matter field causing the string breaking
phenomenon.  Here we want to demonstrate that ({\it i}) the
monopole part of the potential indeed incorporates the effect of
string breaking and ({\it ii}) that it is monopole pairing which
is the reason for the breakdown of the monopole confinement mechanism.
We are aware of
the incompleteness of the analogy to QCD and the relative simplicity of
monopole dynamics in 3 instead of 4 dimensions.

It seems that there is only one possibility to explain string
breaking in three space--time dimensions. We assume that, in the
presence of matter fields, monopoles are increasingly bound
into neutral pairs (magnetic dipoles). The size of a typical pair
should be of the order of the string breaking distance
$R_{\mathrm{br}}$. Indeed, if the distance $R$ between the test
charges is much larger than $R_{\mathrm{br}}$ then the test
charges do not recognize individual monopoles inside the dipoles
(in other words, the fields of the monopoles from the same
magnetic dipole effectively screen each other) and the vacuum is
basically composed of neutral particles. Therefore, at large
inter--particle separations there should be no string tension.
However, if $R \ll R_{\mathrm{br}}$ then the test charges do
recognize individual monopoles even if they are bound in dipoles, and the
monopole fields may induce a piecewise linearly rising potential.
These simple considerations can be made more rigorous by
analytical calculations~\cite{DipoleGas} for a gas of
infinitely small--sized dipoles.

Recently, it was found that the matter fields in the Abelian Higgs
model lead to a logarithmic attraction between monopoles and
anti--monopoles~\cite{KleinertNogueiraSudbo} which results in the
formation of monopole--anti--monopole bound states and string breaking.
The formation of dipoles can also be explained as due to the existence
of Abrikosov--Nielsen--Olesen vortices~\cite{ANO}, the string tension
of which gets increased as we move in the parameter space deeper into
the Higgs region~\footnote{Note that the division of the parameter space
of the model into Higgs and confinement regions is only loose
since these regions -- as we discuss below -- are analytically
connected.}.
Massless quarks also force the Abelian monopoles to
form bound states~\cite{Agasian:2001an}.
Note that the origin of monopole binding in the zero
temperature case of cAHM$_3$ is physically different from the monopole
binding observed at the finite temperature phase transition in compact
QED~\cite{Binding,CISPapers12}. It is different as well from the $Z_2$ vortex
mechanism in the Georgi--Glashow model~\cite{DunneKoganKovnerTekin}.

In this paper we numerically establish a relation between string
breaking on one hand and the occurrence of monopole--antimonopole
bound states on the other by studying some properties of the
monopole ensembles provided by the compact Abelian Higgs model. In
Section~\ref{sec:model} we recall the definition of the model and
discuss its missing ordinary phase transition. In
Section~\ref{sec:potential} flattening of the potential is
described. Here we also introduce the $\eta$ angle as a parameter
which defines the ''effectiveness'' of string breaking.
Section~\ref{sec:cluster} is devoted to an investigation of the
cluster structure of the monopole ensembles. Our conclusions are
presented in the last Section.

\section{The Model and Its Crossover}
\label{sec:model}

We consider the $3D$ Abelian gauge model with a compact gauge field
$\theta_{x,\mu}$ and a Higgs field $\Phi_x$ with unit electric
charge. The coupling between the gauge and the Higgs fields is
$S_{x,\mu} \propto \Re{\mathrm{e}} \,(\Phi^\dagger_{x} e^{i
\theta_{x,\mu}} \Phi_{x+\hat{\mu}})$. To simplify calculations we
consider the London limit of the model, which corresponds to an
infinitely deep potential on the Higgs field. In this limit the
radial part of the Higgs field, $|\Phi_x|$, is frozen and the only
dynamical variable is the phase $\varphi_x$ of this field, $\Phi_x
= |\Phi_x|\, e^{i \varphi_x}$. Thus the Higgs-gauge coupling
reduces to the simple interaction $S_{x,\mu} \propto
\cos(\varphi_{x+\hat{\mu}} - \varphi_{x} + \theta_{x,\mu})$.
However,
the model can be simplified even further by fixing the unitary gauge,
$\varphi_x = 0$ leading to $S_{x,\mu} \propto \cos\theta_{x,\mu}$.
Thus we consider the model with the action
\beqn
S[\theta] = - \beta \sum_P \cos\theta_P  - \kappa \sum_l \cos\theta_l\,,
\label{eq:action}
\eeqn
where $\beta$ is the gauge (Wilson) coupling, $\kappa$ is the hopping
parameter and $\theta_P$ is the plaquette angle.
We study the model at zero temperatures on
lattices of size $L^3$, with $L=12,16,24,32$.

The phase structure of the model on the boundaries of the phase
diagram in the $\beta$--$\kappa$ plane can be established using
the following simple arguments. At zero value of the hopping
parameter $\kappa$ the model (\ref{eq:action}) reduces to the pure
compact Abelian gauge theory which is known to be confining at any
coupling $\beta$ due to the presence of the monopole
plasma~\cite{Polyakov}. This argument extends to the
low--$\kappa$ region of the phase diagram. Therefore we call this
the ''confinement region''. At large values of $\kappa$ (also
called the ''Higgs region'') the monopoles should disappear
because the gauge field in this limit is increasingly restricted
to the trivial vacuum state: $\theta_{x,\mu} = 0$.

At large $\beta$ the model reduces to the three dimensional $XY$
model which is known to have a second order phase transition at
$\kappa^{XY}_c \approx 0.453$~\cite{XYphase}.  Indeed, in this
limit we get the condition $\dd \theta_l  \equiv \theta_P = 0$
which forces the gauge field to be a gauge transformation of the
vacuum, $\theta_{x,\mu} = - \phi_{x+\hat\mu} + \phi_{x} + 2 \pi
l_{x,\mu} \in (-\pi,\pi]$, $l_{x,\mu} \in \Z$, $\phi_{x} \in
(-\pi,\pi]$. The scalar fields $\phi$ are the spin fields in that
model.

Despite the phase structure on the boundary of the coupling plane
is well established, the structure of its interior is still under
debate. Indeed, in Ref.~\cite{NagaosaLee} arguments were given
that the interior is trivial ($i.e.$, there is no ordinary phase
transition for finite values of $\beta$ and $\kappa$) while the
$XY$--phase transition takes place in an isolated point at
$\beta=\infty$. In Ref.~\cite{AnotherOpinion} it has been
suggested that the phase diagram of cAHM$_3$ resembles the
vapor--liquid diagram with a critical end--point. Finally, in
Ref.~\cite{EinhornSavit} it was argued that the phase diagram
contains a ''pocket'' in which a Coulomb phase could be realized.
Arguments given in Ref.~\cite{FradkinShenker} do not allow to
distinguish between these three possibilities.

In a numerical study on rather small
lattices~\cite{BhanotFreedman} no hint for an ordinary phase
transition at finite coupling constant $\beta$ has been found.
However, for simulations allowing fluctuating Higgs lengths,
sufficiently away from the London limit, the phase diagram has
been seen to become nontrivial~\cite{Munehisa}. Recently, the
phase structure of the cAHM$_3$ has been studied by the authors of
Ref.~\cite{Kajantie} in connection with the nature of the
transition in the type-I and the type-II region.
The alleged second order transition in the type-II region away from
the London limit
still remained inconclusive.

\begin{figure}[!htb]
\begin{center}
\begin{center}
\begin{tabular}{cc}
\epsfxsize=6.5cm \epsffile{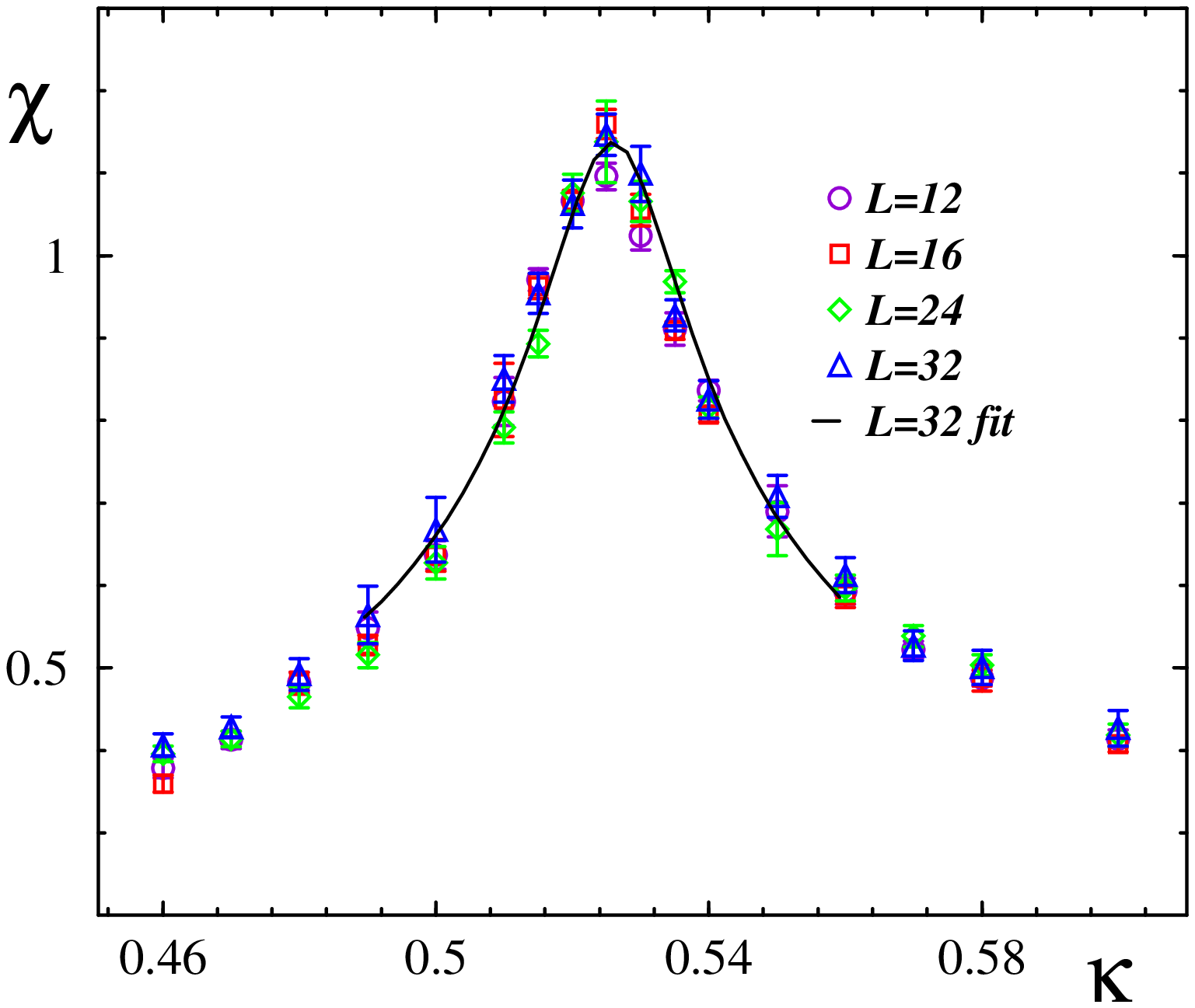} &
\epsfxsize=7.0cm \epsffile{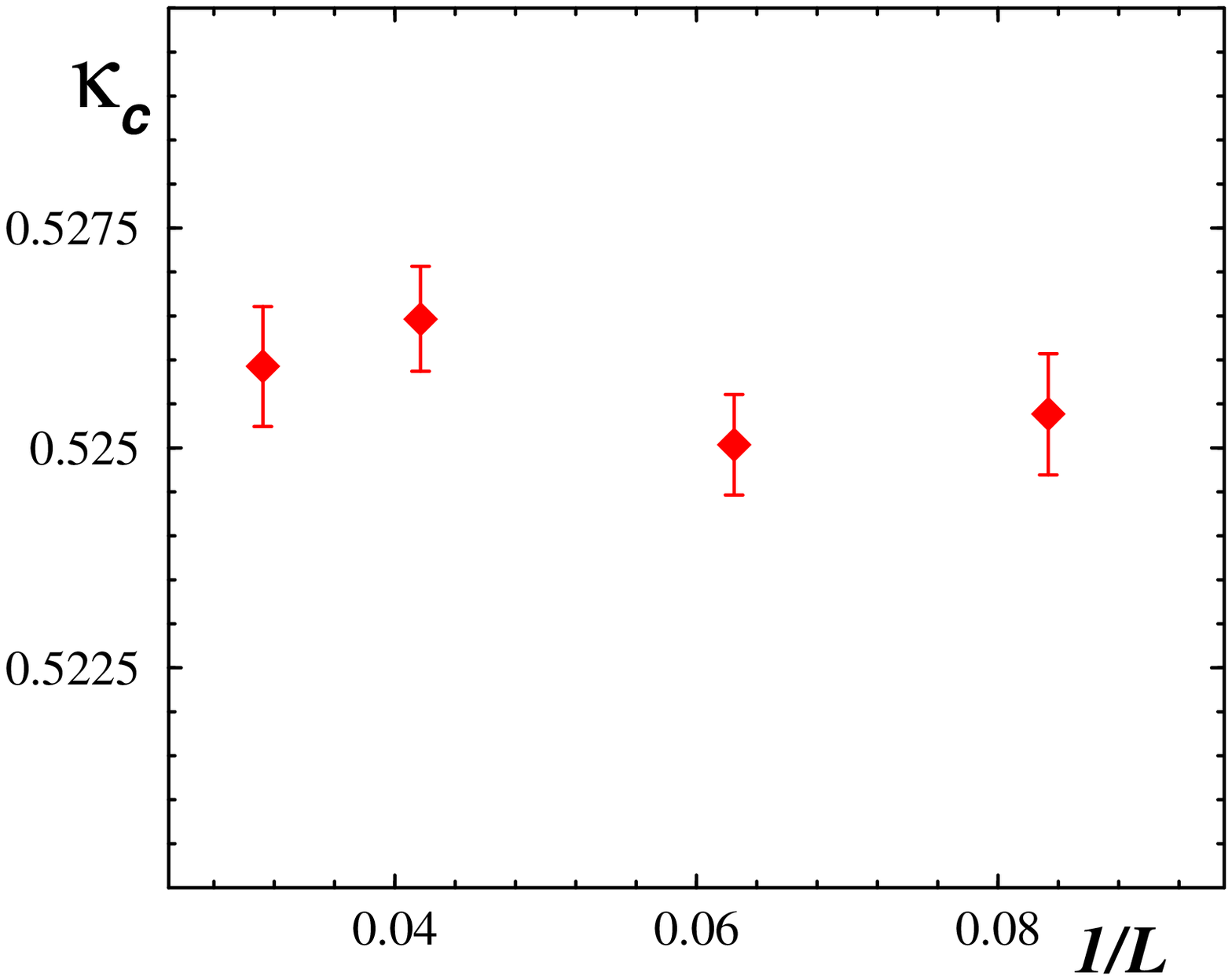} \\
(a) & (b) \\
\end{tabular}
\end{center}
\end{center}
\caption{(a) The susceptibility of the hopping term (\ref{eq:susceptibility})
as a function of $\kappa$;
(b) the crossover point $\kappa_c$ as a function of the inverse lattice size,
$L^{-1}$.}
\label{fig:phase}
\end{figure}

Here we are not going to study the whole phase diagram of cAHM$_3$
although this question would be still interesting. As we describe
below, we observed that at moderately small $\beta$ the Higgs and
confinement regions are connected analytically by a crossover as
predicted in Ref.~\cite{FradkinShenker}. We concentrate on the
changing role of monopoles under the aspect of the string breaking
phenomenon accompanying the crossover at relatively small $\beta$
with increasing hopping parameter $\kappa$. For the simulations we
use a Monte Carlo algorithm similar to the one described in
Ref.~\cite{CISPapers12} and have considered $5 \cdot 10^3$ to $5
\cdot 10^4$ independent configurations per data point, depending
on the lattice size and the set of coupling constants.
We vary the value of the hopping
parameter $\kappa$ at a fixed value of gauge coupling constant
$\beta=2.0$. To locate a (pseudo--) critical point we use the
susceptibility of the hopping term,
\beqn
\chi = \langle S^2_H[\theta]\rangle - {\langle
S_H[\theta]\rangle}^2\,, \quad
S_H[\theta] = - \sum_l \cos\theta_l\,,
\label{eq:susceptibility}
\eeqn
which is shown~\footnote{Note that all figures in this paper are shown for
$\beta = 2.0$.}
in Figure~\ref{fig:phase}(a) for $L=12,16,24,32$.
The height of the peak is practically independent on the lattice size.
We have observed a very similar volume independence also of the
susceptibility of the gauge term, $S_G[\theta] = - \sum_P \cos\theta_P$.
Thus, in agreement with Ref.~\cite{BhanotFreedman}
we conclude that there is {\it no ordinary phase transition}
between the Higgs and confinement regions of the parameter space of the
model.

The crossover point $\kappa_c(L)$ is located fitting the susceptibility
(\ref{eq:susceptibility}) in the vicinity of the peak by the following
function:
\beqn
\chi^{\mathrm{fit}}(\beta) = \frac{C_1}{{[C^2_2 + {(\kappa - \kappa_c)}^2]}^\alpha}\,,
\eeqn
where $C_{1,2}$, $\kappa_c$ and the power $\alpha$ are fitting
parameters. In Figure~\ref{fig:phase}(a) we show the fit of the
susceptibility data for the $32^2$ lattice.
The fit parameters practically do not depend on the lattice
size. We depict the critical value of the hopping parameter
$\kappa_c$ $vs.$ the inverse lattice size $L^{-1}$ in
Figure~\ref{fig:phase}(b). The value of the power $\alpha$ is very
close to $1 \slash 4$. In the next Sections we will work with the
lattice $32^2$ which passes the crossover at $\kappa_c =
0.526(1)$ along a line with fixed $\beta=2.0$.

\section{The Flattening of the Potential}
\label{sec:potential}

String breaking manifests itself in the flattening of the
potential between test particles with (opposite) electric charges
$q=\pm 1$. In principle, we can separate the contributions to the
potential from monopoles {\it and} from the rest (''photon
contribution''). Monopoles are responsible for the string tension.
Therefore one can expect that the monopole contribution alone will
signal the onset of string breaking when the monopole dynamics
starts changing. It would be much more demanding to extract the string
part from the full potential and to study its change over the
parameter space of the model. The full potential contains also the
perturbative photon contribution which -- being logarithmically
large at small distances -- shadows the eventually linearly rising
part. Any statement about the string part would require a careful
fit of full potential. On the more technical side, the monopole
contribution alone, calculated separately according to the
configurations generated in the simulation of the AHM, has a much better
signal/noise ratio compared to the full potential. All this
justifies to proceed directly to the evaluation of the monopole
contributions to the external--charge potential.

To this end we have divided the gauge field  $\theta_l$ into a regular
(photon) part and a singular (monopole) part~\cite{PhMon}:
\beqn
  \theta = \theta^{\phot} + \theta^{\mon}\,, \quad
  \theta^{\mon} = 2 \pi \Delta^{-1}_3 \delta p[j]\,.
\label{eq:theta-decomp}
\eeqn
The 0-form $\dual j \in \Z$ is nonvanishing on the sites dual to the
lattice cubes $c$ which are occupied by monopoles~\cite{DGT}:
\beqn
  j_c = \frac{1}{2\pi} \sum\limits_{P \in \partial c} {(-1)}^P \,
  {[\theta_P]}_{\mathrm{mod} \, 2 \pi} \; ,
\label{jc}
\eeqn
where the factor ${(-1)}^P$ takes the plaquette orientations
relative to the boundary of the cube into account.
In Eq. (\ref{eq:theta-decomp}) the
2-form  $p_P[j] = [\theta_P]$ (the notation $[\cdots]$
means taking the integer part) corresponds to the Dirac strings
living on the links of the dual lattice, which are either closed or
connecting monopoles with
anti--monopoles, $\delta \dual p[j] = \dual j$.
While $\dual j$ is gauge invariant, the 2-form  $p_P[j]$ is not.
For the Monte Carlo configurations provided by the simulation of
(\ref{eq:action}) we have located the Dirac strings, $p[j]\ne0$, and
constructed the monopole part $\theta^{\mon}$ of the gauge field
according to the last equation in (\ref{eq:theta-decomp}).
The operator $\Delta^{-1}_3$ in Eq.~(\ref{eq:theta-decomp})
is the inverse lattice Laplacian defined for a three--dimensional
lattice $L^3$:
\beqn
  \Delta^{-1}_d (\vec x;L) = \frac{1}{2 L^d}
  \sum\limits_{{\vec p}^2 \neq 0}  \frac{e^{i (\vec p,\vec x)}}
  {d - \sum^d_{i=1} \cos p_i}\,,
\eeqn
where $p_i = 2 \pi k_i \slash L_i$ for $k_i = 0,\dots, L_i - 1$,
with $i=1,\dots, d$ and $L_i=L$.

We define the potential between test particles with the help
of the following correlator of two Polyakov loops:
\beqn
{\langle P({\vec 0}) P^\dagger ({\vec R})\rangle} = e^{- L V(R)} \; ,
\label{eq:polyakov}
\eeqn
located at two--dimensional points ${\vec 0}$ and ${\vec R}$. The potential
$V$ depends on $R=|{\vec R}|$. The use the Polyakov loop has clear
advantages compared to the Wilson loops. The construction of
the Polyakov loops is not only possible for finite--temperature
but also for finite--volume cases. $L=L_i$
is the common length of the zero--temperature box in all three
directions. Due to the absence of space--like links joining the
Polyakov loops the correlator~(\ref{eq:polyakov}) defines the
static component of the potential.
Note that the monopole contribution to the Polyakov loop
correlator (\ref{eq:polyakov}) does not depend on the precise form of
the Dirac string $\dual p[j]$. Therefore this contribution
is gauge--invariant.

We discuss the results for the potential using the following fitting
function:
\beqn
e^{- L V^{\mathrm{fit}}(R)} = C_0 \Bigl[\sin^2 \eta
+ \cos^2 \eta\,
\frac{\cosh(\sigma L (L/2 - R))}{\cosh(\sigma L^2/2)} \Bigr] \cdot
\exp\Bigl\{ \gamma L \Bigl[\Delta^{-1}_{2}(R) - \Delta^{-1}_{2}(0)\Bigr]
\Bigr\}\,,
\label{eq:fit-potential}
\eeqn
where  $C_0$, $\eta$, $\sigma$ and $\gamma$ are fitting parameters and
$\Delta^{-1}_2$ is the inverse lattice Laplacian in two dimensions.

The meaning of the expression (\ref{eq:fit-potential}) is quite simple.
In the absence of
string breaking and in an infinite two--dimensional volume the
leading contribution to the function in the right hand side of Eq.
(\ref{eq:fit-potential}) should be just $\const \cdot e^{- \sigma
L R}$ where $\sigma$ is the
effective string tension.
Here ''effective'' means that this term gives rise
to a linear part in the potential at short distances.

The string breaking manifests itself in the appearance of an
additional constant term, $const_1 + const_2 \cdot e^{- \sigma L
R}$. Next, the finiteness of the two--dimensional volume reduces
the exponential to the $\cosh$--function
which takes care of the symmetry
$R \to L - R$. Finally, we introduced a Coulomb term in order to
take into account sub-leading corrections.

The dimensionless parameter $\eta \in [0,\pi \slash 2]$ -- which
we call a ''breaking angle'' -- has a sense only as long as
$\sigma \neq 0$. It can be considered as a kind of ''order
parameter'' for string breaking: if $\eta=0$, no string breaking
occurs, and if $\eta = \pi \slash 2$, the potential does not
contain a linear piece at all. An intermediate value of the
breaking angle implies the existence of the finite distance
$R_{\mathrm{sb}}$ at which the string between the test particles
breaks.
Note that we have introduced a normalizing $\cosh$--factor in the
second term in the brackets in order to keep the
$V^{\mathrm{fit}}(R=0)$ value independent on $\eta$. This
definition is a matter of conventions.

To justify the presence of the Coulomb--like term in the fitting
function~(\ref{eq:fit-potential}) let us consider three dimensional
compact QED. It is well known that in the Villain representation
the Polyakov loop correlator factorizes into the photon and
monopole contribution. The monopole contribution can be evaluated
exactly and it contains a massless pole, $\Delta^{-1}_2(R)$,
corresponding to the Coulomb potential between test particles. The
total correlator should not contain the massless pole due to the
massiveness of the photon. Therefore the monopole contribution to
the correlator must contain -- in addition to the linear term --
the difference between the Yukawa and Coulomb potentials,
$\Delta^{-1}_2(R;m) - \Delta^{-1}_2(R;m=0)$ corresponding to the
exchange by ''real'' (massive) and ''bare'' (massless) photons. Here
$\Delta^{-1}_2(R;m)$ is the propagator of a particle with the mass
$m$. The mentioned above sub-leading term is small at
distances smaller than the inverse photon mass. However, this term
gives a significant (logarithmically growing) contribution at
larger separations between test particles. Thus the largest
deviation from the linear behaviour of the monopole contribution
to the potential is expected to come from large distances due
to exchange of a massless (bare) photon.

Similar arguments should apply to the case of the compact AHM. The
bare photon here, however, is not massless due to the spontaneous
breaking of the $U(1)$ symmetry. Therefore the fitting
function~(\ref{eq:fit-potential}) should be modified: the Coulomb
potential should be replaced by the Yukawa one. We have found that
such fits do not work well because the corresponding Yukawa mass
turns out to be consistent with zero within huge error bars. On
the other hand, the mass of the bare photon should be small at the
QED side of the crossover where the form of the
fit~(\ref{eq:fit-potential}) is obviously justified. We have found
numerically that this fitting function works well also at the
Higgs side of the crossover. Therefore in Eq. (\ref{eq:fit-potential})
we restrict ourselves to the Coulomb term only.
\begin{figure}[!htb]
\begin{center}
\begin{center}
\begin{tabular}{cc}
\epsfxsize=6.8cm \epsffile{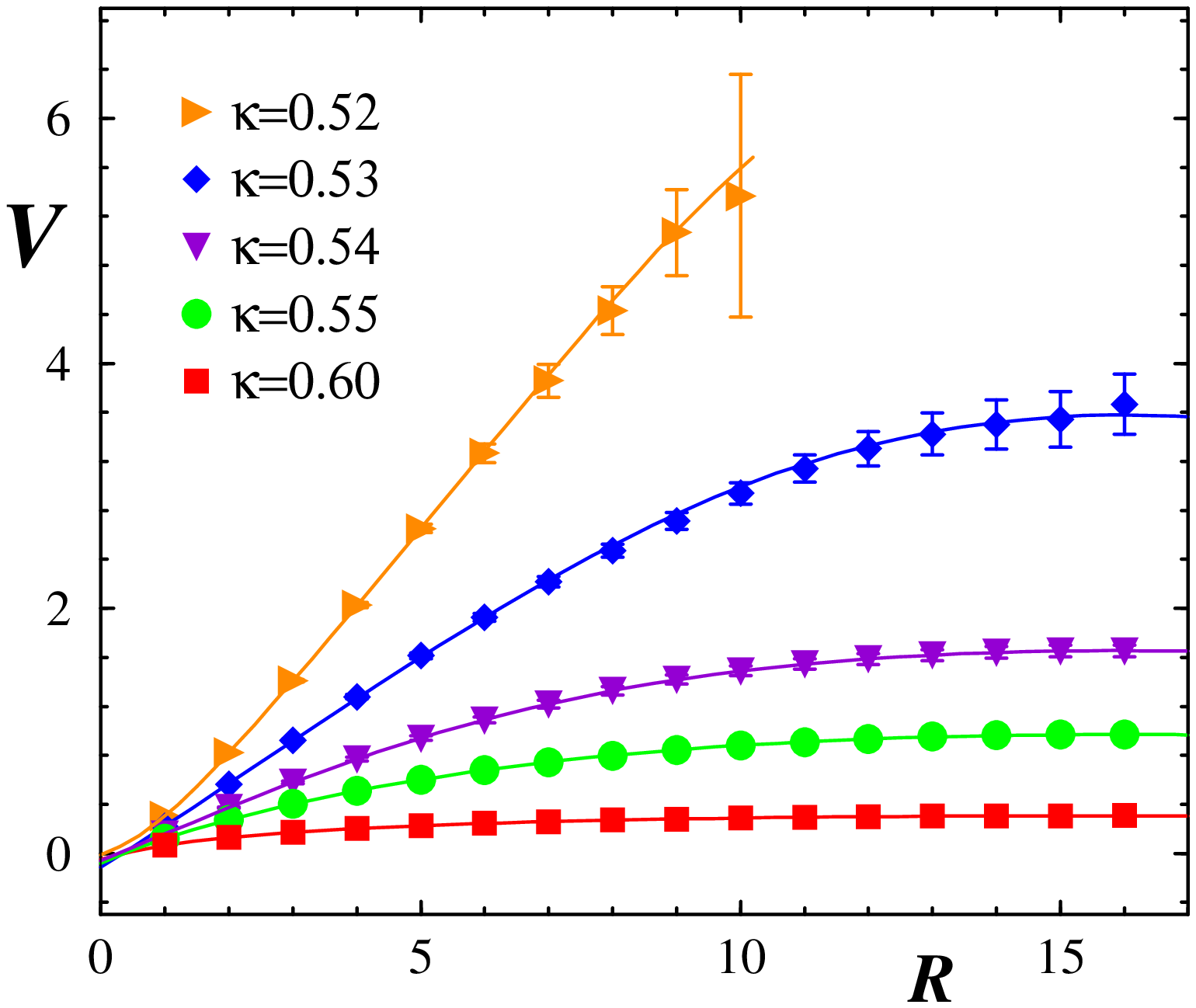} &
\epsfxsize=7.0cm \epsffile{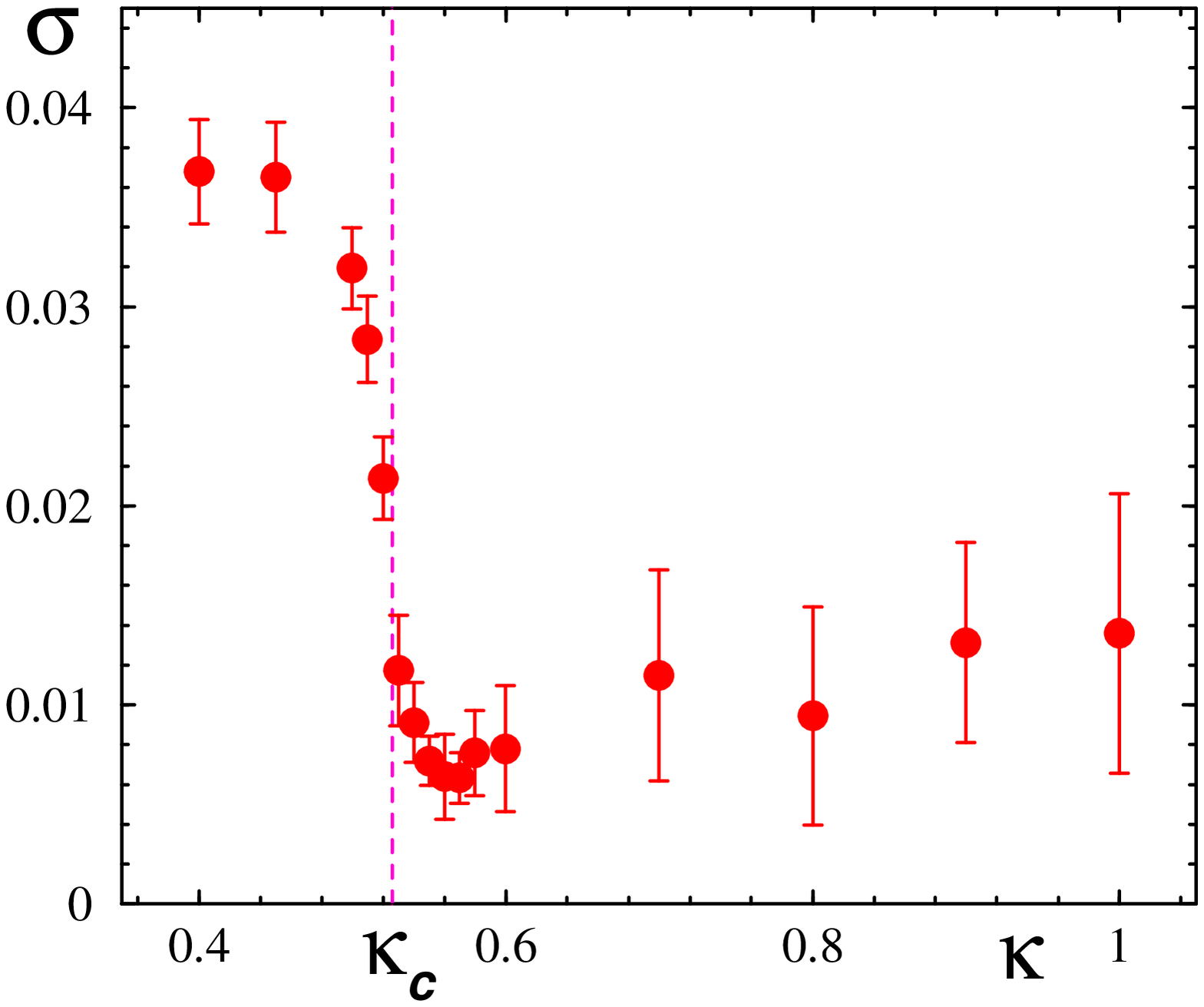} \\
(a) & (b) \\
\end{tabular}
\end{center}
\end{center}
\caption{(a) The potential for
$\kappa=0.52$,
$\kappa=0.53$,
$\kappa=0.54$,
$\kappa=0.55$ and
$\kappa=0.60$ extracted from the monopole contribution to the Polyakov
loop correlator by Eq. (\ref{eq:polyakov}). The fits by the function
(\ref{eq:fit-potential}) are shown by solid lines.
(b) The string tension {\it vs.} $\kappa$. In this and all subsequent
figures the string breaking transition at $\kappa_c$ for $\beta=2.0$
is marked by a vertical line.}
\label{fig:potential}
\end{figure}

The fits of the numerical data for the potential $V(R)$ due to monopoles
by the expression (\ref{eq:fit-potential}) are shown in
Figure~\ref{fig:potential}(a) for five values of the hopping
parameter from $\kappa=0.52$ (below string breaking)
to $\kappa = 0.60 $ (far from the transition on the Higgs side)
including $\kappa=0.53 \approx \kappa_c$ (in the vicinity of the
transition).
In the fits of the potential the point $R=0$ was excluded.
One can clearly recognize a linear part in the potential near the
transition point. As $\kappa$ increases (this corresponds to
moving deeper into the Higgs region) the linear part gradually
disappears. This can also be seen from the properties of the
string tension $\sigma$ shown in Figure~\ref{fig:potential}(b). The string
tension itself, which on the confinement side
amounts roughly to 50 \% of the QED$_3$ string tension
(corresponding to $\kappa=0$),
drops to a smaller value over a very narrow $\kappa$ region.
The described behaviour of the potential is consistent with the
expected disappearance of isolated monopoles on the Higgs side of
the string breaking transition. The residual string tension, which
is accompanied by a short string breaking length $R_{\mathrm{sb}}$,
can be accounted for by the monopole--antimonopole dipoles of
finite size. With $\eta \rightarrow \pi/2$  the fit error of
$\sigma$ increases.

\begin{figure}[!htb]
\begin{center}
\begin{center}
\begin{tabular}{cc}
\epsfxsize=6.7cm \epsffile{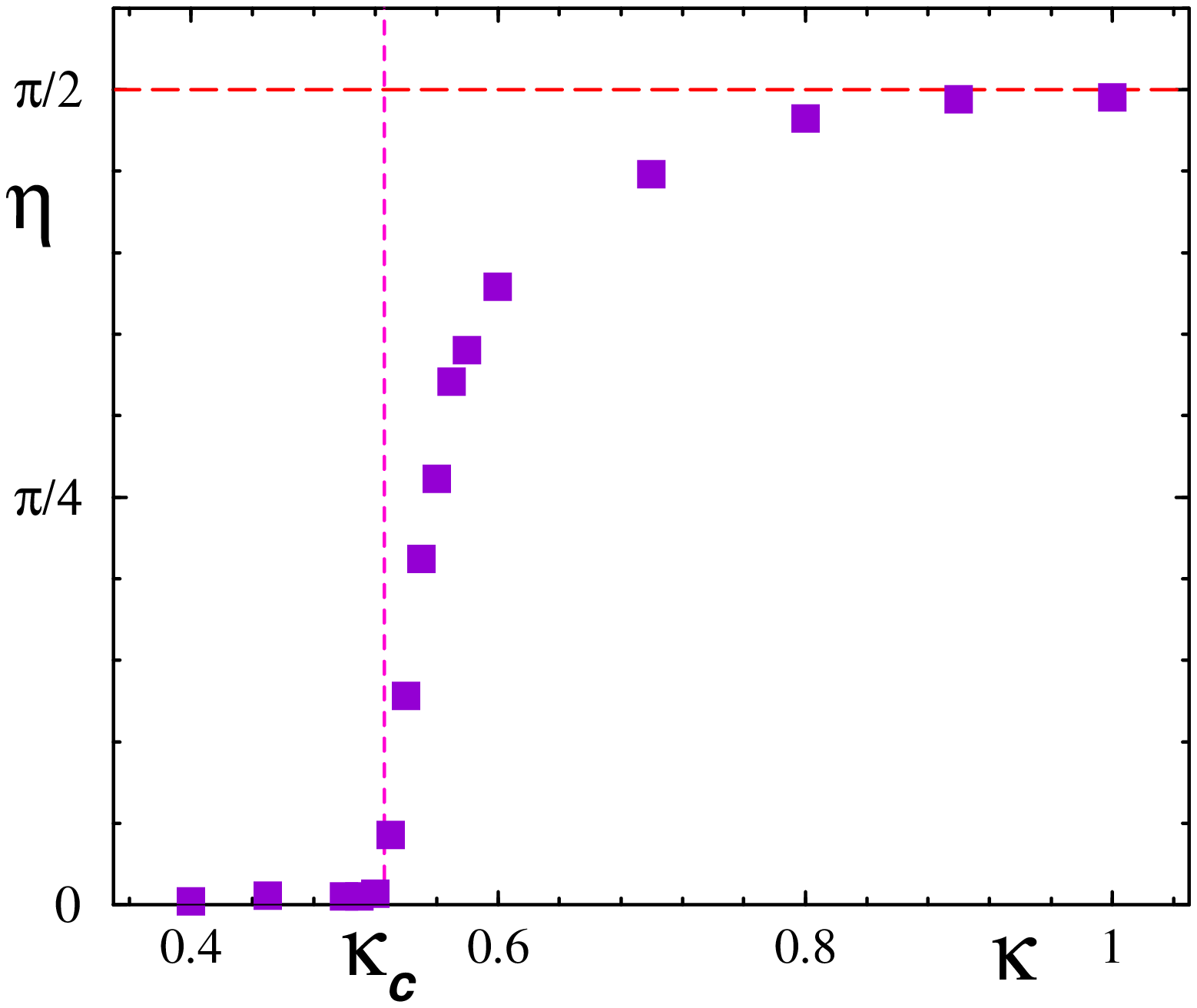} &
\epsfxsize=7.0cm \epsffile{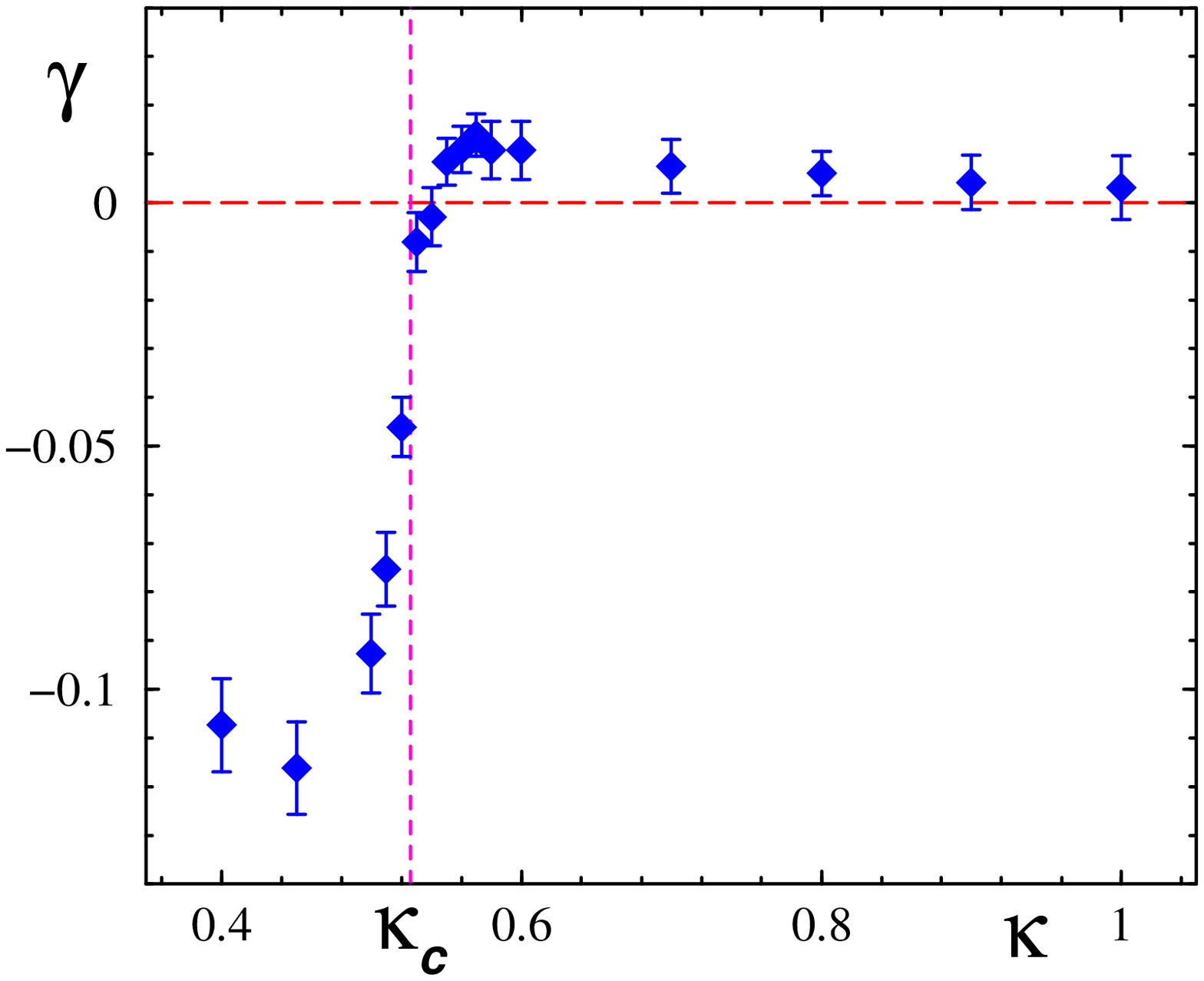} \\
(a) & (b) \\
\end{tabular}
\end{center}
\end{center}
\caption{
(a) The breaking angle $\eta$ and
(b) the parameter $\gamma$ appearing in the fitting
function~(\ref{eq:fit-potential}) {\it vs.}
the hopping parameter $\kappa$.}
\label{fig:potential:aux}
\end{figure}
The breaking angle $\eta$ is shown in
Figure~\ref{fig:potential:aux}(a) as a function of $\kappa$.
It clearly shows an ''order-parameter--like''
behaviour: it is close to zero for $\kappa < \kappa_c$ and it is
finite at $\kappa > \kappa_c$. Small values of $\eta$ imply that
the string breaking distance is still large. At $\kappa \sim 1$
the value of $\eta \sim \pi \slash 2$ indicates that the area--law
term in the Polyakov loop correlator~\eq{eq:fit-potential} has
become irrelevant.

The parameter $\gamma$, shown in
Figure~\ref{fig:potential:aux}(b), seems to vanish
on the Higgs side of the string breaking transition. This may indicate that in
the Higgs region the ''bare'' photon mass becomes significant and
that the corrections to the linear potential gets concentrated at
small distances~\footnote{We remind the reader that the smallest
distance, $R=0$, is excluded from the fit.}. Thus, long distance
corrections should be zero, {\it i.e.} $\gamma \sim 0$.

\section{The Cluster Structure of the Monopole Ensemble}
\label{sec:cluster}

In this section we turn to the monopole clustering aspect of the
Monte--Carlo configurations which have been used in the last section
to work out the monopole part of the external--charge potential.
We closely follow Ref.~\cite{CISPapers12} where the cluster analysis
of the monopole configurations in the case of compact QED$_3$  at
non--zero temperature was performed.

\begin{figure}[!htb]
\begin{center}
\begin{tabular}{cc}
\epsfxsize=7.0cm \epsffile{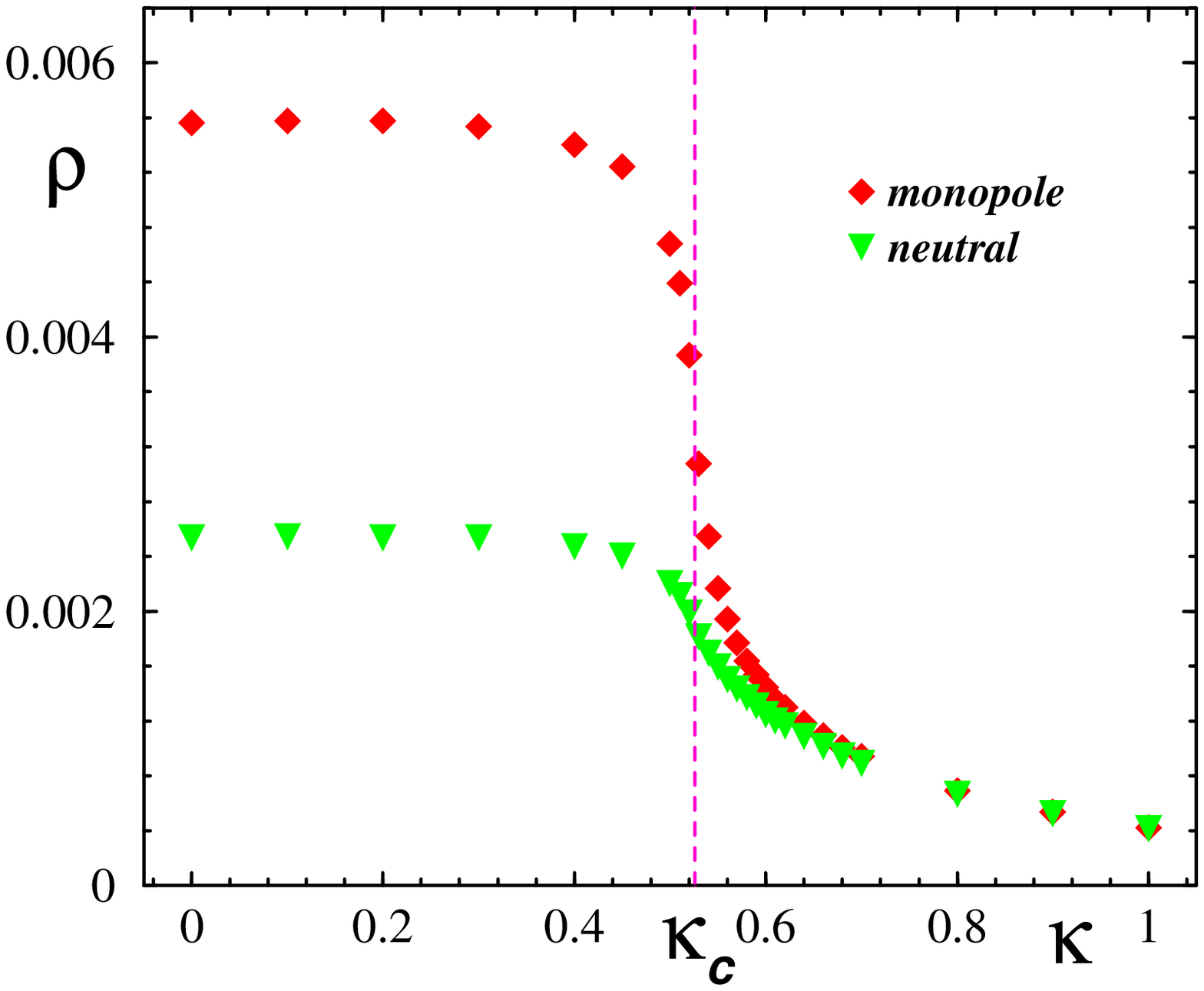} &
\epsfxsize=6.7cm \epsffile{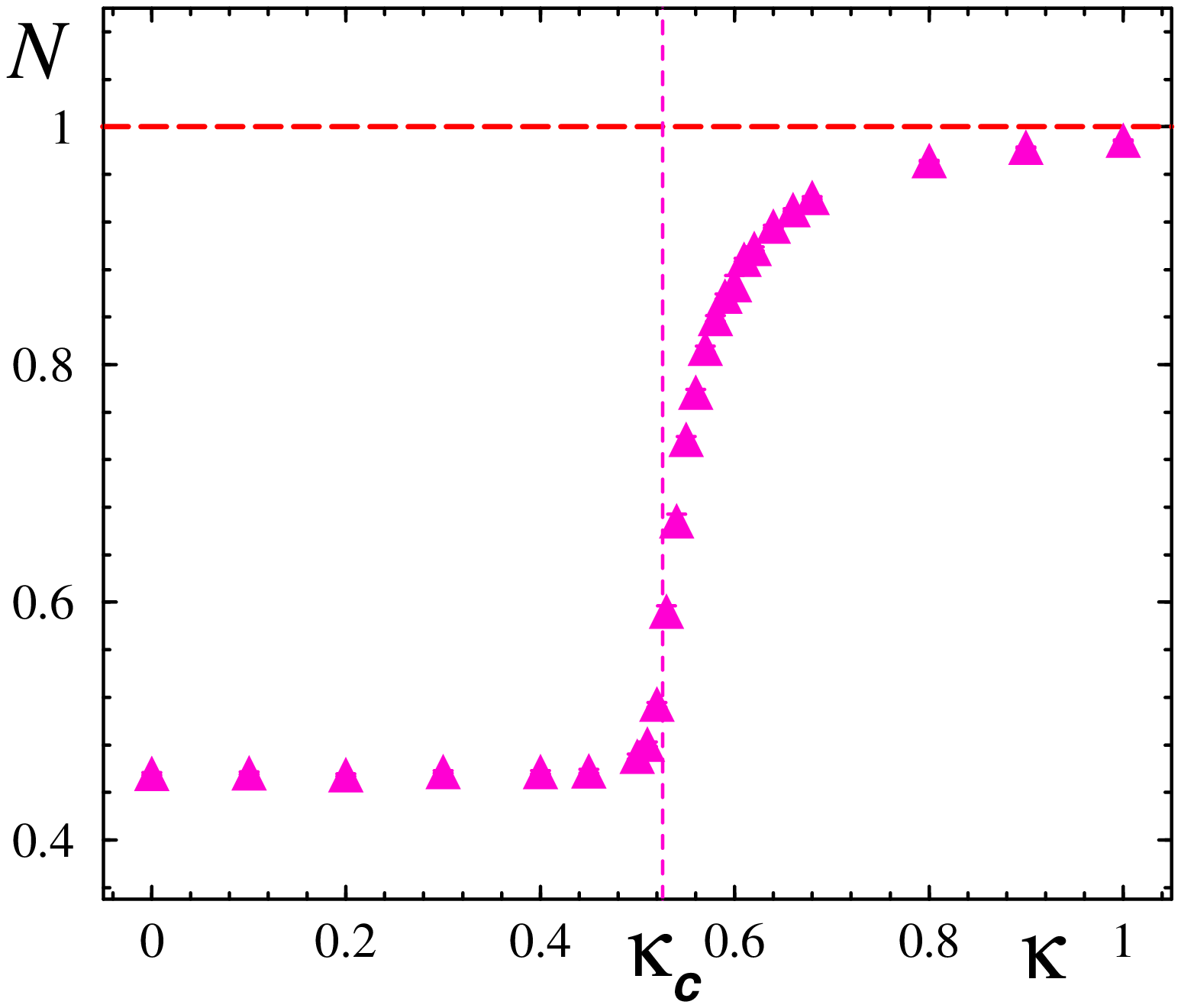} \\
(a) & (b)
\end{tabular}
\end{center}
\vskip -4mm
\caption{(a) Total density of monopoles (diamonds) and
density of the monopoles in neutral clusters (triangles), and (b)
the fraction of neutral clusters among all clusters, both as functions
of $\kappa$.}
\label{fig:cluster}
\end{figure}

The simplest quantity describing the behaviour of the
monopoles is the monopole density,
$\rho = \sum_c |j_c|/L^3$, where $j_c$ is
the integer valued monopole charge inside the cube $c$ defined
in Eq.~\eq{jc}. The density of the total number of monopoles is a
decreasing function of the hopping parameter $\kappa$ as it is
shown in Figure~\ref{fig:cluster}(a) by diamonds.
The density sharply drops down at $\kappa_c$, which has been
recognized as the string breaking transition point, but the
density does not vanish on the Higgs side of the crossover. The
binding of monopoles into dipoles should show up as an increase of
the number of monopoles enclosed in neutral clusters. We call a
monopole cluster neutral if the charges of the corresponding
constituent monopoles sum up to zero. Clusters are connected
groups of monopoles and anti--monopoles where each object is
separated from at least one neighbor belonging to the same cluster
by a distance less or equal than some $R_{\mathrm{max}}$.
The smallest clusters are isolated (anti-)monopoles. In our
analysis we have used $R^2_{\mathrm{max}}=3~a^2$ which means that
monopoles are considered as neighbors if their cubes
share at least one single corner.

We show also in Figure~\ref{fig:cluster}(a), symbolized by triangles,
the density of monopoles in neutral clusters which
almost covers the total density on the Higgs side
of the string breaking transition.
If we take into account that also bigger dipoles
-- which cannot be identified by our procedure -- may be formed,
this clearly signals the binding transition.

In an alternative, perhaps more clear way this is illustrated by the
fraction of monopoles belonging to neutral clusters,
$N= \rho_{\mathrm{neutral}} \slash \rho_{\mathrm{total}}$, which is shown in
Figure~\ref{fig:cluster}(b).
Being constant on the confinement side of the string breaking transition,
this quantity starts suddenly to rise at the transition.
This indicates that at the transition point (crossover)
the binding process rapidly takes place. At large $\kappa$ the fraction is
very close to unity. Then all monopoles are bound.

\begin{figure}[!htb]
\begin{center}
\begin{tabular}{cc}
\epsfxsize=7.0cm \epsffile{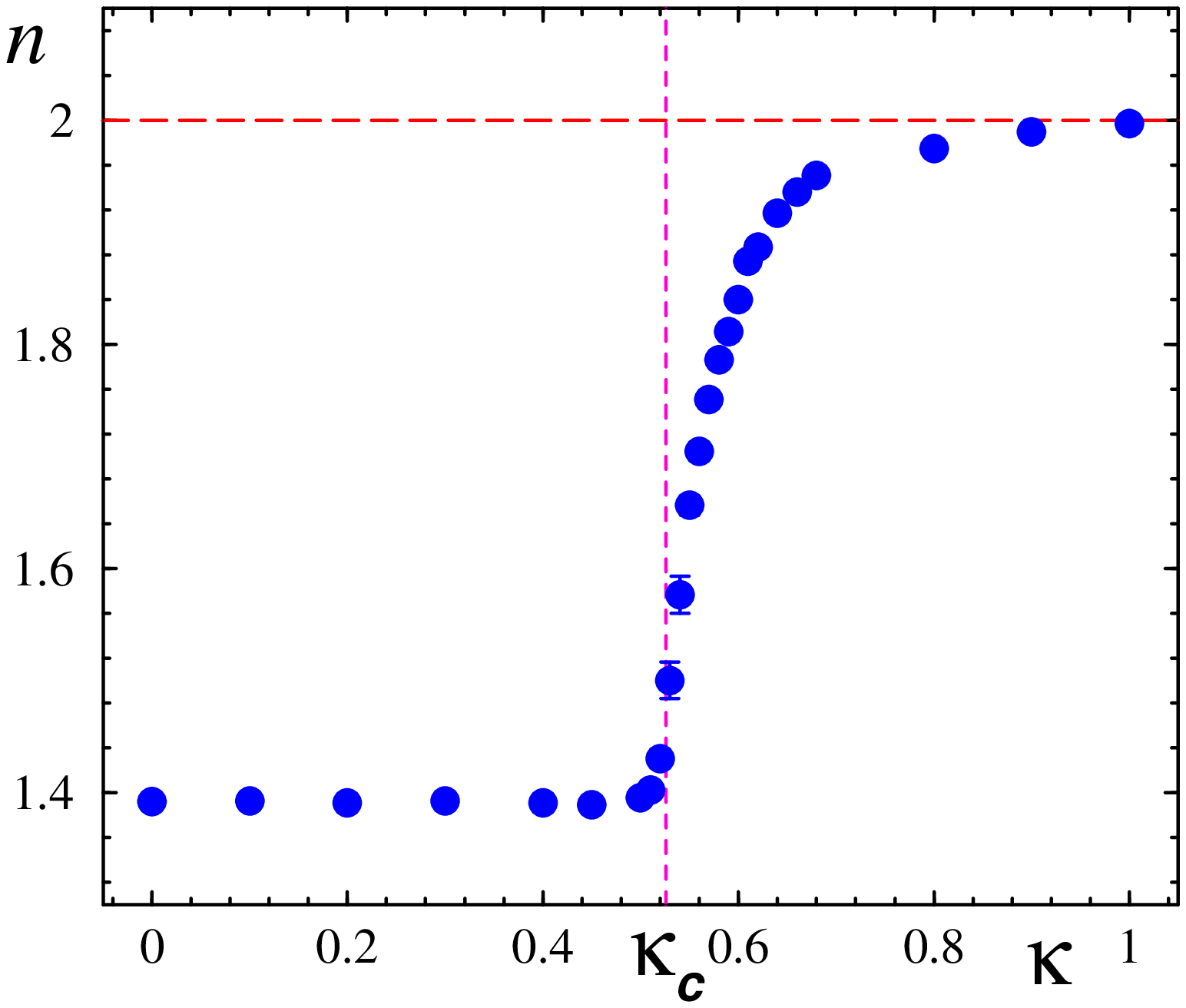} &
\epsfxsize=6.8cm \epsffile{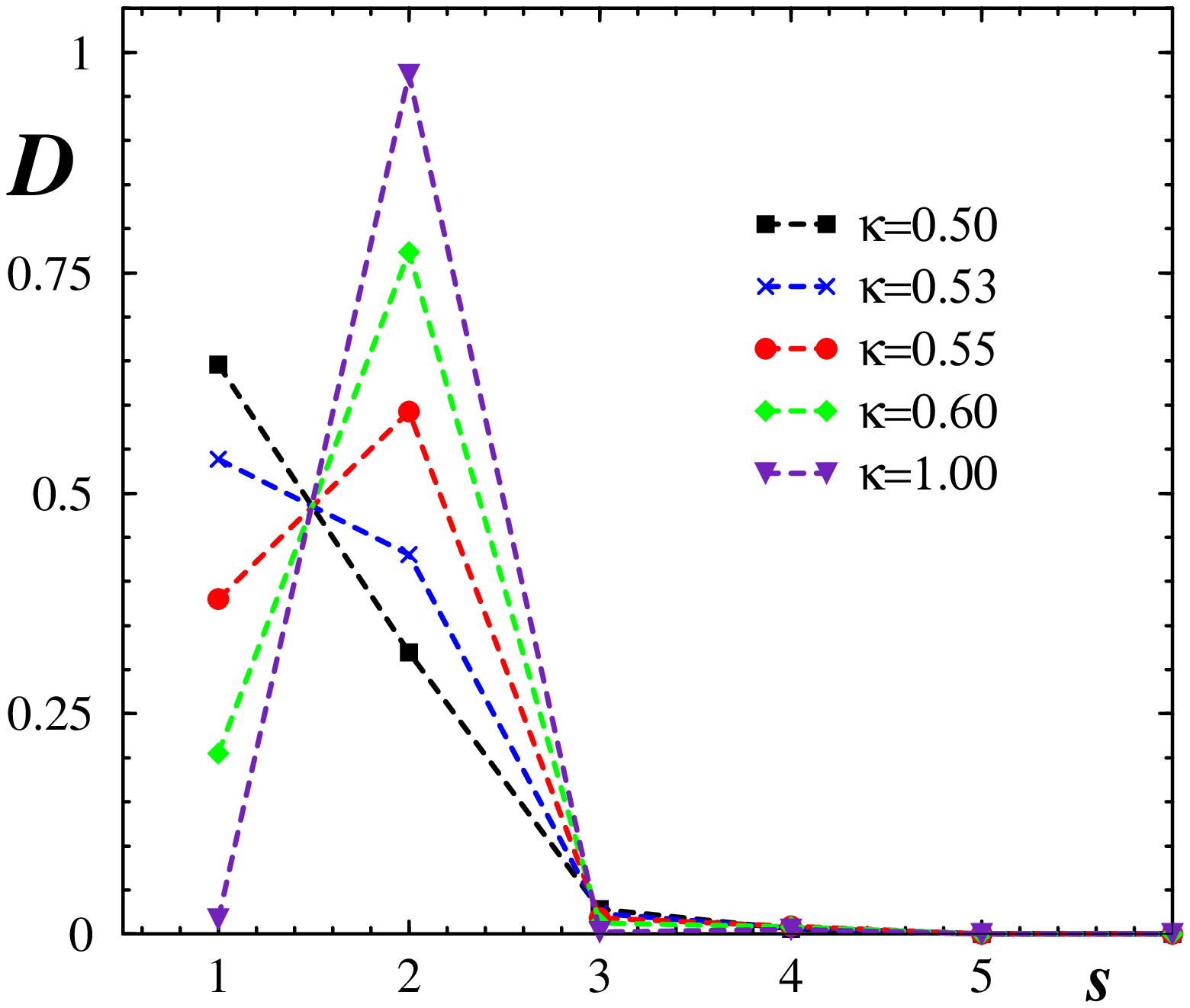} \\
(a) & (b)
\end{tabular}
\end{center}
\vskip -4mm
\caption{(a) The average number of monopoles and
anti-monopoles per cluster as function of $\kappa$ and (b) the
(normalized) cluster size distribution $D(s)$ for a few values of
$\kappa$.} \label{fig:cluster:aux}
\end{figure}
Finally, in Figure~\ref{fig:cluster:aux}(a) we present the average number
of (anti-)monopoles per cluster and in Figure~\ref{fig:cluster:aux}(b)
the (normalized) cluster size distribution $D(s)$ where $s$ is the number
of (anti-) monopoles in the cluster,
for a few values of the hopping parameter $\kappa$. On the
confinement side of the string breaking transition
($\kappa \approx 0.5$) the vacuum consists
to $\approx 70$ \% of {\it isolated monopoles}. At the crossover
(the string breaking transition) at $\kappa \approx 0.53$
the number of isolated monopoles decreases,
and on the Higgs side ($\kappa > 0.53$) the vacuum is dominated
by the dipole gas.

\section{Conclusions}

We have numerically observed that in the London limit
of the three--dimensional Abelian Higgs model string breaking
occurs and is accompanied by monopole recombination into dipoles, in
agreement with arguments given in Ref.~\cite{KleinertNogueiraSudbo}.

Our study shows that the monopole binding is not necessarily
accompanied by an ordinary phase transition of first or second
order. There is a proposition~\cite{KleinertNogueiraSudbo},
however, that the string breaking may be associated with a
Berezinsky--Kosterlitz--Thouless type transition due to the
appearance of an anomalous dimension of the gauge field induced by
the fluctuations of the matter fields. This possibility is not
ruled out by our results. In the London limit (studied in this
article) the fluctuations of the radial components of the matter
field are suppressed, while far away from the London limit the
fluctuations become significant such that an ordinary phase
transition may exist~\cite{Munehisa,Kajantie}.

\section*{Acknowledgments}
The authors are grateful to V.~G.~Bornyakov for valuable
discussions and A.~Sudb\o \ for critical comments. M.~N.~Ch. is
supported by the JSPS Fellowship P01023. E.-M.~I. gratefully
appreciates the support by the Ministry of Education, Culture and
Science of Japan (Monbu-Kagaku-sho) and the hospitality extended
to him by H. Toki at RCNP.

\end{document}